*PERSPECTIVE*

# Deep Learning in Pharmacogenomics:
# From Gene Regulation to Patient Stratification

Alexandr A. Kalinin[1,2,*], Gerald A. Higgins[1,*], Narathip Reamaroon[1], S.M. Reza Soroushmehr[1], Ari Allyn-Feuer[1], Ivo D. Dinov[1,2,4], Kayvan Najarian[1,3], Brian D. Athey[1,4,5,6,§]

[1]Department of Computational Medicine and Bioinformatics, University of Michigan Medical School, Ann Arbor, MI, USA
[2]Statistics Online Computational Resource (SOCR), University of Michigan School of Nursing, Ann Arbor, MI, USA
[3]Department of Emergency Medicine, University of Michigan Medical School, Ann Arbor, MI, USA
[4]Michigan Institute for Data Science (MIDAS), University of Michigan, Ann Arbor, MI, USA
[5]Department of Internal Medicine, University of Michigan Health System, Ann Arbor, MI, USA
[6]Department of Psychiatry, University of Michigan Medical School, Ann Arbor, MI, USA

[*]Equal contribution
[§]Corresponding author

## Abstract

This *Perspective* provides examples of current and future applications of deep learning in pharmacogenomics, including: (1) identification of novel regulatory variants located in non-coding domains of the genome and their function as applied to pharmacoepigenomics; (2) patient stratification from medical records; and (3) the mechanistic prediction of drug response, targets, and their interactions. Deep learning encapsulates a family of machine learning algorithms that has transformed many important subfields of artificial intelligence (AI) over the last decade, and has demonstrated breakthrough performance improvements on a wide range of tasks in biomedicine. We anticipate that in the future, deep learning will be widely used to predict personalized drug response and optimize medication selection and dosing, using knowledge extracted from large and complex molecular, epidemiological, clinical, and demographic datasets.

# 1 Introduction

Machine learning (ML) is a fundamental concept of Artificial Intelligence (AI), and is a key component of the ongoing Big Data revolution that is transforming biomedicine and healthcare [1-3]. Unlike many "expert system"-based methods in medicine that rely on sets of predefined rules about the domain, machine learning algorithms learn these rules from data, benefiting directly from the detail contained in large, complex, and heterogeneous datasets [4]. Deep learning is one of the most successful types of machine learning techniques that has transformed many important subfields of AI over the last decade. Examples include data modeling and analytics, computer vision, speech recognition, and natural language processing (NLP). Deep learning demonstrated breakthrough performance improvements over pre-existing techniques on a wide range of complex tasks across multiple biomedical research domains spanning from basic clinical to translational [5].

The deep learning methods landscape encompasses a variety of biologically-inspired models that can be applied directly to raw data, automatically learn useful features, and make a prediction without a need to form a hypothesis [5]. While biomedical applications of deep learning are still emerging, they have already shown promising advances over the prior state-of-the-art in several tasks [6-8]. We anticipate deep learning algorithms to have a substantial impact on pharmacogenomics, pharmaceutical discovery, and, more generally, on personalized clinical decision support in the near future.

Pharmacogenomics (PGx) focuses on the identification of genetic variants that are correlated with drug effects in populations, cohorts, and individual patients. It has traditionally straddled the intersection of genomics and pharmacology, with the greatest impact on clinical practice in oncology [9], psychiatry [10], neurology [11], and cardiology [12]. Pharmacogenomics offers promise for applications such as medication optimization for patients based on genotype in diagnostic testing, value as a companion diagnostic (CDx), and drug discovery and development. However, physicians, caregivers, patients, and pharmaceutical and biotechnology companies have all been slow to adopt pharmacogenomics, despite recommendations by the U.S. Food and Drug Administration (FDA) [13]. Recently, however, pharmaceutical companies that are faced with rising costs and resource investments required for drug development, have begun to recognize the potential of genomics for drug discovery, and to a lesser extent, for stratification of participants in clinical trials to mitigate adverse events and increase efficacy [14, 15]. In addition, the adoption of pharmacogenomic testing for optimization of medication selection in psychiatry has shown promise for the clinical utility [10, 16].

Historically, the exome has provided a rich source of single variants for genotyping in pharmacogenomics, with a focus on genes that encode ADME (absorption, distribution, metabolism, excretion), including DMET (drug metabolizing enzymes and transporters) proteins. More recently, the non-coding regulatory genome is proving to be the next domain for the discovery of new pharmacogenomic variants that will provide clinical utility [17]. This new research lies at the heart of the new field of "Pharmacoepigenomics", which is the corresponding emerging subdomain of pharmacogenomics that focuses on studying the role of the epigenome in drug response [18]. However, these new variant discovery methods and their potential corresponding drug development processes can be very time consuming, where large trials are needed to assess clinical efficacy, toxicity, and safety [19]. Moreover, the continuing growth of other types of collected data that can improve phenotype-driven therapy via pharmacogenomics also poses a number of challenges for accurate treatment response and outcome prediction (**Figure 1**).

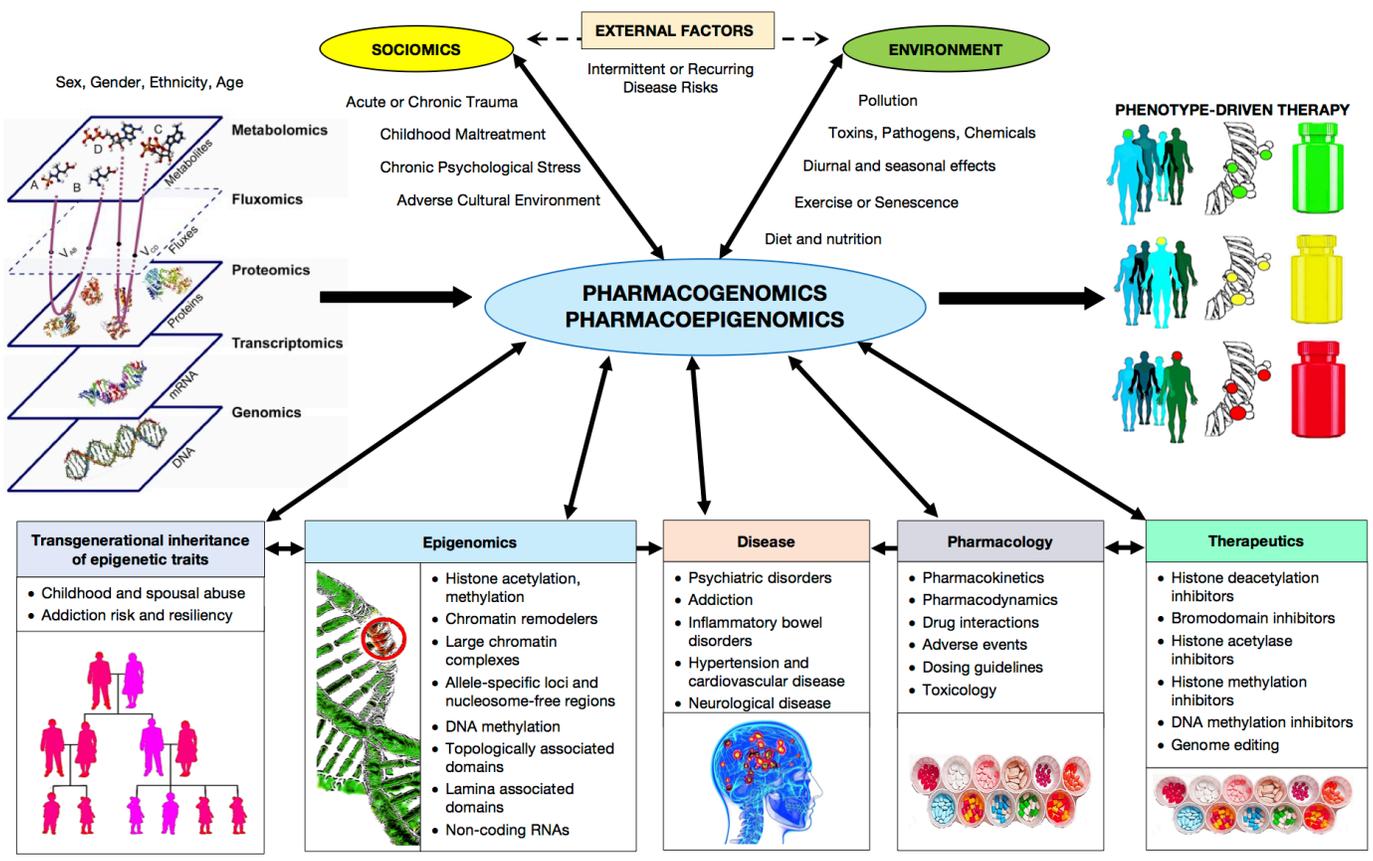

**Figure 1. Pharmacogenomics and pharmacoepigenomics for phenotype-driven therapy.** Pharmacogenomics and pharmacoepigenomics will be facilitated by various data sources that include not only traditional "omics" databases, but also: growing knowledge in epigenomics, including novel variant and their functional annotation; pharmacological data, including new therapeutics, drug interactions, and dosing guidelines; and patient data, including clinical, socio-economic, and information about the environment. Partially adapted from [20] under CC BY 3.0 license.

Extracting usable knowledge from large databases requires advanced computational methods that can find patterns, conduct prediction, detection, and classification [2, 3] along with visual data analytics [21, 22]. Current approaches for knowledge extraction in pharmacogenomics include statistical methods [23, 24], machine learning [24, 25], and, recently, deep learning. Therefore, new deep learning-based predictive analytic methods are desirable to accelerate the discovery of new pharmacogenomic markers, forecast drug efficacy in stratified cohorts of patients to minimize potential adverse effects of drugs, and to maximize the success of treatment.

In this *Perspective,* we provide examples of current and potential future applications of AI, and more specifically deep learning, in pharmacogenomics and pharmacoepigenomics to illustrate the utility and the future potential of these methods. It should be recognized that this *Perspective* provides an overview of selective pharmacogenomic applications, and is not intended to provide an authoritative and rigorous evaluation of the technical foundations of these methods. In addition, like other machine learning methods, deep learning techniques are prone to error if not grounded in computational, statistical, and experimental expertise [6]. Indeed, proper controls and performance metrics are critical for the performance evaluation of such models [26]. For those seeking extensive reviews, there are a number of existing deep learning application-focused review articles in computational biology [6, 7, 27], pharmacology and drug discovery [28-32], and other areas of biomedicine [8]. The most successful applications of deep learning require expertise in both the methodology and in the subject matter under investigation: in this case, pharmacogenomics. Thus, we have two synergistic objectives—to increase awareness while stimulating dialogue among pharmacogenomics and pharmacology researchers about promising future applications of this powerful machine learning computational methodology.

## 2 Opportunities and challenges for deep learning applications

**2.1 Methodological advantages of deep learning**

Conventional machine learning algorithms are typically limited in their ability to process raw data [5]. Their performance heavily depends on the extraction of relevant representations or features that requires careful engineering and considerable domain expertise (**Figure 2A**). In the past, biomedical datasets have typically been limited by sample size, and since often many more features could be measured, the performance of conventional machine learning algorithms degraded when useful information was buried in an excess of extracted features. This posed a challenge for the determination and extraction of the optimal feature set for the problem under examination. Two related and widely-used solutions are used to overcome this limitation: 1) dimensionality reduction methods that shrink the feature space to the set of most informative components [24]; and 2) feature selection methods that identify a relatively small number of features that can accurately predict an outcome [25]. While many of these general-purpose methods already exist, they are not necessarily optimized for pharmacogenomic biomarker discovery. This and other related pharmacological research applications require careful experimental design and choice of validation techniques. Overall, limitations of conventional machine learning methods include the need for extensive human guidance, painstaking feature handcrafting, careful data pre-processing, and the above-mentioned dimensionality reduction to achieve top performance.

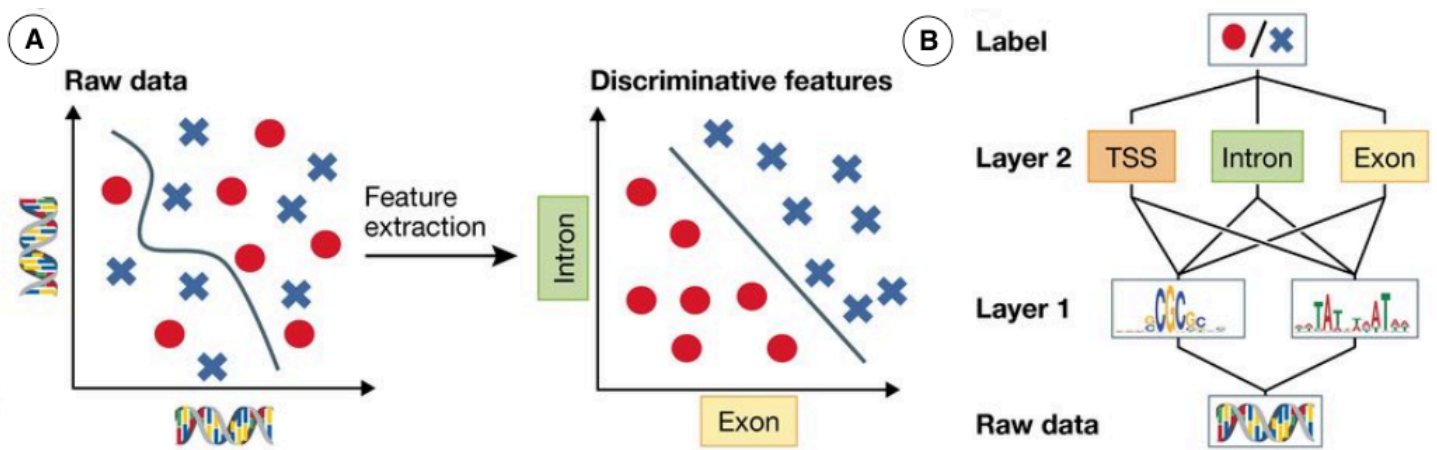

**Figure 2. Feature extraction vs representation learning.** (A) Raw input data are often high-dimensional and related to the corresponding label in a complicated way, which is challenging for many conventional machine learning algorithms (left plot). Alternatively, higher-level features extracted using a deep model may be able to better discriminate between classes (right plot). (B) Deep networks use a hierarchical structure to learn increasingly abstract feature representations from the raw data recommendation. Adapted from [7] under CC BY 3.0 license.

In contrast, deep learning methods model data by learning high-level representations with multi-layer computational models such as artificial neural networks (ANNs) [5]. While classic feed-forward ANNs might serve as drop-in replacement for other machine learning models and require input pre-processing and feature extraction; deep learning architectures, such as convolutional neural networks (CNNs), allow the algorithm to automatically learn features from raw and noisy data. Deep neural networks rely on algorithms that optimize feature engineering processes to provide the classifier with relevant information that maximizes its performance with respect to the final task. Such deep learning models can be thought of as automated "feature learning" or "feature detection," which facilitates learning of hierarchical, increasingly abstract representations of high-dimensional heterogeneous data [5], also known as "representation learning" (**Figure 2B**). Some common deep learning methods include deep feed-forward artificial neural networks (ANNs), convolutional neural networks (CNNs), recurrent neural networks (RNNs), stacked auto-encoders (SAEs), deep belief networks (DBNs), and deep reinforcement learning techniques [5-7, 27]. In biomedicine, these models are capable of unguided extraction of highly complex, nonlinear dependencies from raw data such as raw sequence data [8].

Recent applications of deep learning in biomedicine have already demonstrated their superior performance compared to other machine learning approaches in many biomedical problems [8], including those in image analysis [33-36], genomics [7, 27]; as well as drug discovery and repurposing [30, 31]. This great success of deep learning models in many tasks is thought to be enabled by the explosive growth of volume of raw data along with significant progress in computing, including the use of powerful graphical processing units (GPUs) that are specifically well-suited for the optimization of deep learning models.

**Figure 3** shows an idealized collective example of deep learning applications in pharmacogenomics. First, deep neural networks are trained on various existing datasets and/or their combinations. Depending on the type of data and a task in hand, prediction outcomes for a dataset can be known (supervised learning), partially known (semi-supervised learning), or not-known-at-all (unsupervised learning). Due to the flexibility of architectures, neural networks are capable of multimodal learning, i.e. jointly learning from several different datasets and data types without explicit definition of common features [37]. For example, Chaudhary et al. [38] trained a deep auto-encoder model jointly on RNA-seq, miRNA-seq, and methylation data from The Cancer Genome Atlas (TCGA) to predict sub-groups of hepatocellular carcinoma patients. Moreover, deep networks can be used in a multi-task learning (MTL) regime by learning multiple objectives simultaneously and providing several outputs such as prediction of the regulatory function of a sequence, pathway mapping, disease and ADE mark identification, drug efficacy, and dosage recommendation [31].

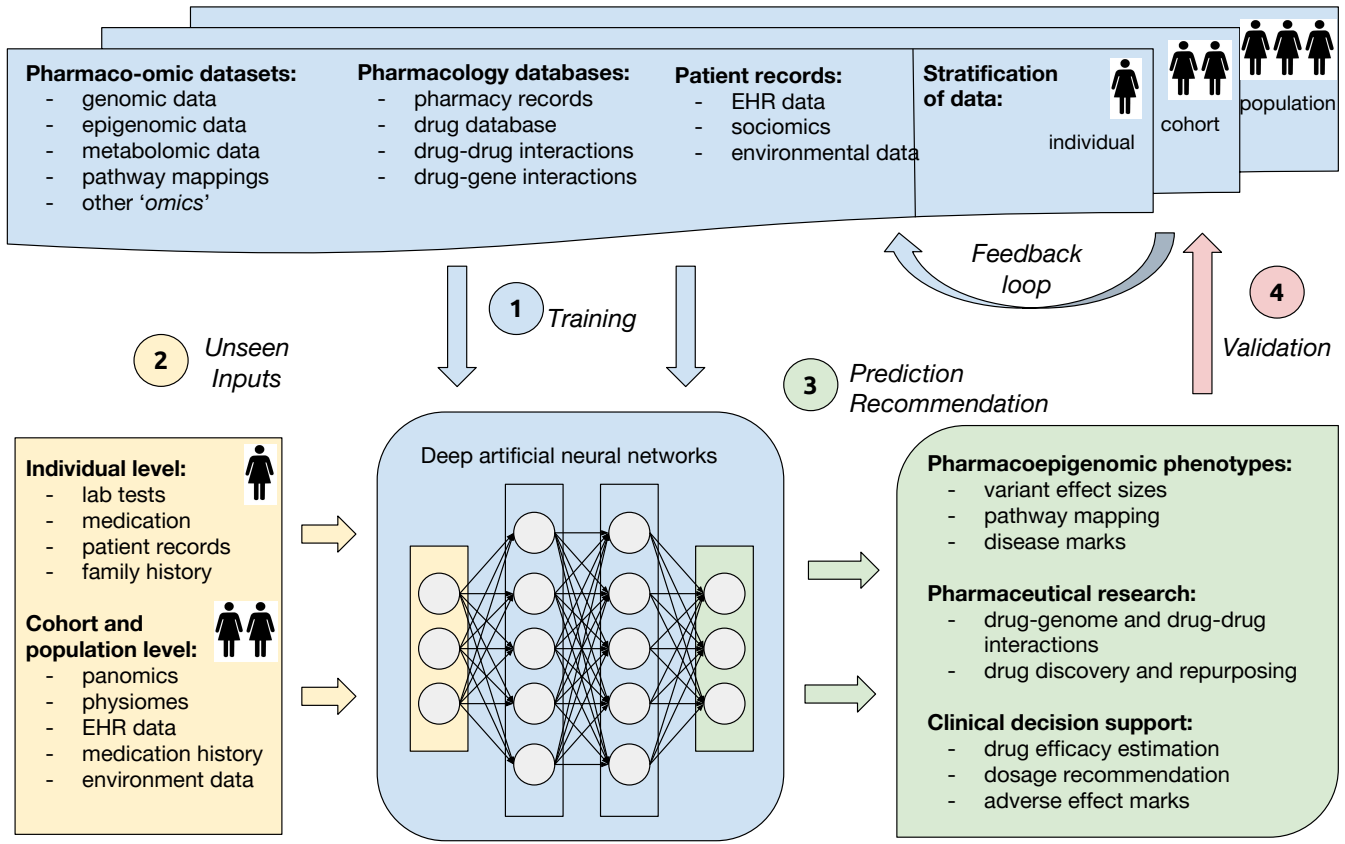

**Figure 3. An idealized example of deep learning applications to a number of common problems in pharmacogenomics, including prediction of pharmacoepigenomic phenotypes, novel regulatory variants and their function, and stratified clinical decision support.** (1) First, deep neural networks are trained on various existing datasets and/or their combinations with known outcomes. (2) To perform prediction, unseen data of the same format as training samples are given to trained networks as inputs. (3) At the prediction step, a model produces outcome(s) that it was trained to predict during the training phase, for example, personalized drug response, probability of adverse events, or novel pharmacogenomic variants and their pathway mapping. (4) Validation of the predicted responses occurs when true outcomes become available on various population stratification levels. For example, it can be individual patient treatment response, clinical trial results, or new pharmaceutical study results. Data with observed true outcomes can again be used as a training set via a feedback loop, as shown.

## 2.2 Challenges and limitations of deep learning

Deep learning is a very fast-paced research domain and although its potential utility is impressive, there are several challenges associated with the practical usage of these models. Typically, CNNs are complex, heavily parameterized models, with little theoretical guarantees of performance or proven ways to construct effective problem-specific architectures. Due to their complexity, time-consuming training process, and high representational power, such models should be used in cases when sufficient volumes of input data are available and conventional machine learning fails to provide an effective, simpler solution. For biological and clinical applications, adoption of deep learning is slower due to a few reasons, including skepticism arising from the data and hardware requirements, as well as the "black-box" nature of these algorithms, which specifically challenges the notion of model interpretability [5, 8]. Below we discuss some of these challenges in more detail.

**2.2.1 Data requirements**

While most successful deep learning applications relied on large labeled data, many biological and clinical datasets until recently were limited by amount of available labeled samples compared with the big data analytics applications such as natural image processing and NLP. Over time, as the number of samples increases and the number of relevant high-quality labeled datasets expands, the wealth of pertinent pharmacogenomics data that can be used for analytics will begin to resemble a big data challenge on par with contemporary applications in other domains. The rich variety of heterogeneous data types in pharmacogenomics can improve the utilization of highly flexible deep networks that can deal with sparse, high-dimensional, multi-modal data. The real power of deep learning in a domain such as pharmacogenomics will be realized when it is combined with a prior domain knowledge [8], such as gene networks or pathways; a relevant example being used for the prediction of the pharmacological properties of drugs using transcriptomic data combined with pathway information [39]. Multimodal, multi-task, and transfer learning are often used to alleviate data limitations to some extent. Transfer learning approaches include training a deep network on a large existing dataset, and then using this pre-initialized model to learn from a smaller dataset, which typically leads to improved performance [8]. When training data is not (fully) labeled, various semi-supervised techniques can be employed [8, 34, 40]. Data quality is another important concern in deep learning applications. Although deep learning models can be trained directly on raw data, low quality datasets may require additional pre-processing and cleaning. Publicly sharing the pre-processing code (e.g., Basset [41]) and cleaned data (e.g., MoleculeNet [42]) is important to expedite further research and practical applications.

**2.2.2 Overfitting in deep learning**

A trained machine learning model may represent some attributes of the dataset that do not accurately reflect their underlying relationships. This problem may be magnified as the size of the feature or parameter set is increased relative to the size of the input sample set. Such models exhibit poor predictive performance, as they over-represent minor variations in the training data. Overfitting is an issue of trade-off between generalization and approximation of the training data in a model. A model can underfit high-dimensional, heterogeneous dataset with complex hidden patterns if the model's representational power is low, which is often the case, for example, for linear models. Although overfitting is a common issue in machine learning, it is more likely to affect complex models, especially when there are not enough samples in the training set, learning is performed for too long, or where training examples are rare, causing the learner to follow the patterns in training data too closely. In the case of deep learning, overfitting is often a threat due to the high representational power of a model, which leads to the ability to "memorize" the whole training set very efficiently. Thus, deep learning methods require careful choice of model architecture and hyper-parameters. Although there are specific methods to prevent overfitting [5, 7], in general, the trained model should exhibit robust performance on test data using relevant properties of the trained data. For more detailed description of overfitting and model selection, see [43].

Preventing overfitting also requires a very careful design of the model evaluation scheme, including usage of cross-validation techniques, normalization by subjects, etc. Validation metrics may include mean absolute error or root-mean-square error (sample standard deviation of the differences between predicted and observed outcomes) for regression; accuracy; precision (also known as positive predictive value (PPV) – the fraction of retrieved instances that are relevant); recall (sensitivity - the fraction of relevant instances that are retrieved); precision-recall curve (PRC) and area under the PRC (AUPR); receiver operating characteristic (ROC) and area under the ROC curve (AUC); and mean average precision (MAP), for ranking [26, 44]. Although some of these may seem intuitive, correct determination requires great care, and is often fraught with sources of error that are not easily understood, except in the context of the problem under study. For example, while the AUC plot is a common visual method for classification performance evaluation; and it is not the most informative when classes are represented largely by a different number of samples in the dataset [44], which is a common situation in pharmacogenomics, see **Figure 4**. One test of the quality of the trained machine learning model is its ability to faithfully generalize into varying test sets that constitute different manifestations of the same problem.

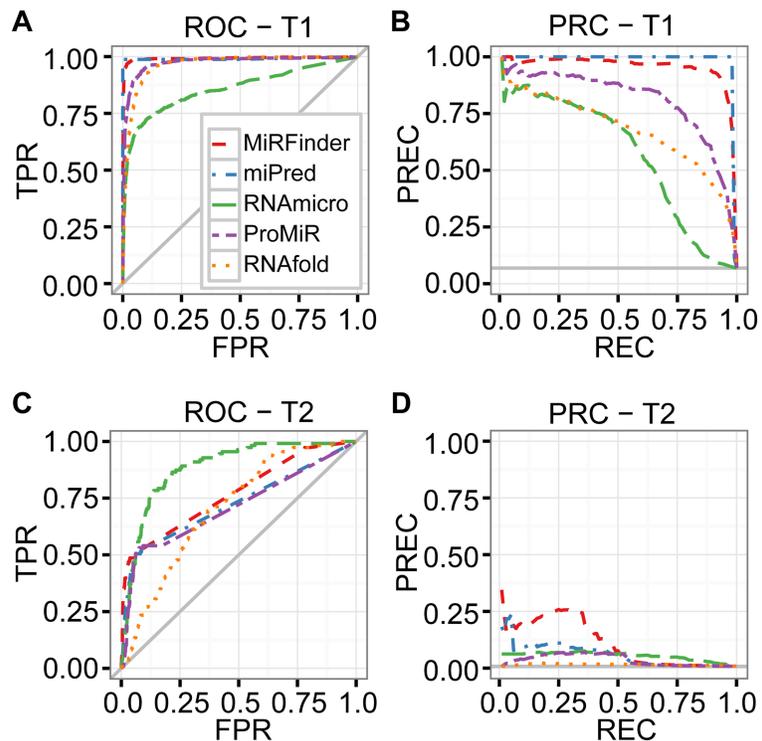

**Figure 4. A re-evaluation of a previously published study confirms the advantages of the PRC plot over the ROC plot in an imbalanced binary classification task.** ROC and PRC plots show the performances of six different tools, MiRFinder (red), miPred (blue), RNAmicro (green), ProMiR (purple), and RNAfold (orange) for microRNA (miRNA) gene discovery from MiRFinder study [45]. The re-analysis used two independent test sets, T1 and T2. The four plots are for (A) ROC on T1, (B) PRC on T1, (C) ROC on T2, and (D) PRC on T2. Adapted from [44] under CC BY 4.0 license.

### 2.2.3 Interpretability of deep learning models

As applications of deep learning models in biomedicine emerge, the question of interpreting a model's outputs receives more attention from the domain experts and practitioners [8]. Unfortunately, the opinion that deep learning models are un-interpretable "black boxes" still prevails outside of machine learning community. However, multiple different methods for deep learning model interpretation have been developed in recent years, including perturbation and back-propagation techniques to evaluate example-specific importance scores, exaggeration of hidden representations, activation maximization, and other methods [6, 8]. In addition, recent studies distinguish between interpreting a model and interpreting its decision, arguing that although interpreting a complex model is hard, often users only want an explanation for the prediction made by the model for a given example [46]. While most applications that motivated model interpretability techniques come from computer vision problems, there is a growing body of research that considers these methods in the life science and biomedical research context.

For example, DeepSEA [47] and DeepBind [48], deep learning-based algorithmic frameworks for predicting the chromatin effects of sequence alterations introduced individual virtual mutations in the input sequence to evaluate the change of the corresponding output (see also **Table 2)**. A similar approach was used by Umarov et al. [49], where the sequence within each sliding window was substituted with a random sequence to estimate its' effect on the result. In order to assess the change in predicted sequence accessibility, the Basset framework [41] implemented insertions of known protein-binding motifs into the centers of input sequences. DeepLIFT [50] allows computing feature importance scores based on explaining the difference of the output from some 'reference' output in terms of differences of the inputs from their 'reference' inputs. Lanchantin et al. [51] applied activation maximization to genomic sequence data to demonstrate patterns learned by an individual neuron in a network. Deming et al. [52] applied the attention mechanism to models trained on genomic sequence. Attention mechanisms provide insight into the model's mechanism of prediction by revealing which portions of the input are used by different outputs. While the interpretability of deep neural networks does not match that of most Bayesian models, recent developments in this area makes it possible to interpret deep learning models, as well as many other commonly used machine learning algorithms such as Support Vector Machines (SVMs) with non-linear kernels or ensemble methods such as Random Forests [8]. For more detailed discussion of interpretability we refer the reader to Ching et al. [8].

# 3 Identification of regulatory pharmacoepigenomic variants, drug target discovery, and patient stratification

**3.1 Inference of pharmacoepigenomic variants: the biology and machine learning**

The non-coding human genome consists of a wealth of elements that regulate gene expression, and accounts for the largest untapped potential source of drug targets and missing pharmacogenomic variation that has yet to be fully exploited [18]. Previous studies [53, 54] demonstrate that psychotropic drugs such as Valproic Acid and Lithium exert their impact in the human CNS through chromatin interaction pathways [54, 55], regulating transcriptional networks that are constrained by the spatial and temporal dimensions of the 4D Nucleome [18, 56]. This led to the recognition that over 90% of the pharmacogenomic SNPs associated with drug efficacy and adverse events in genome-wide association studies (GWAS) are located within enhancers, promoters and intron regions and impact their regulatory function. Coupled with a paucity of coding variation, this certainly contributes to drug response and comprises as yet uncharacterized pharmacogenomic variance requiring further exploration.

Results from the 4D Nucleome program, funded by the U.S. National Institutes of Health, have led to the realization that significant molecular variation, which accounts for human differences in medication response and adverse events are likely based on the intricate organization of the spatial genome [18, 56]. Spatial and temporal morphological changes in the nucleus and nucleoli are associated with the underlying reorganization of the chromatin architecture in 3D and 4D (time dimension added) [57, 58]. Cell type-specific activation of topologically-associated domains (TADs), which are defined areas within chromosome territories (chromosomes) and which provide a foundation for regulation of gene expression, while other nuclear zones of transcriptional regulation include repression (silencing) of gene expression within lamina-associated domains (LADs) at the nuclear periphery **(Figure 5)** [59]. Nuclear pore complexes control the flow of molecular transport, splicing, and are transcriptional hubs. Nucleoli are surrounded by heterochromatin and help to serve as an organizing scaffold for chromosome territories. Inter-chromosomal spatial contacts provide a mechanism for gene regulation between adjacent chromosomes, while chromatin loops within TADs provide one way in which enhancers and promoters regulate gene expression, both in *cis* and in *trans*. The cytoskeleton spans the nucleus via SUN and CASH proteins and may exert drug-induced mechanical control of gene expression. Large transcriptional hubs contain multiple chromatin interaction loops and TADs.

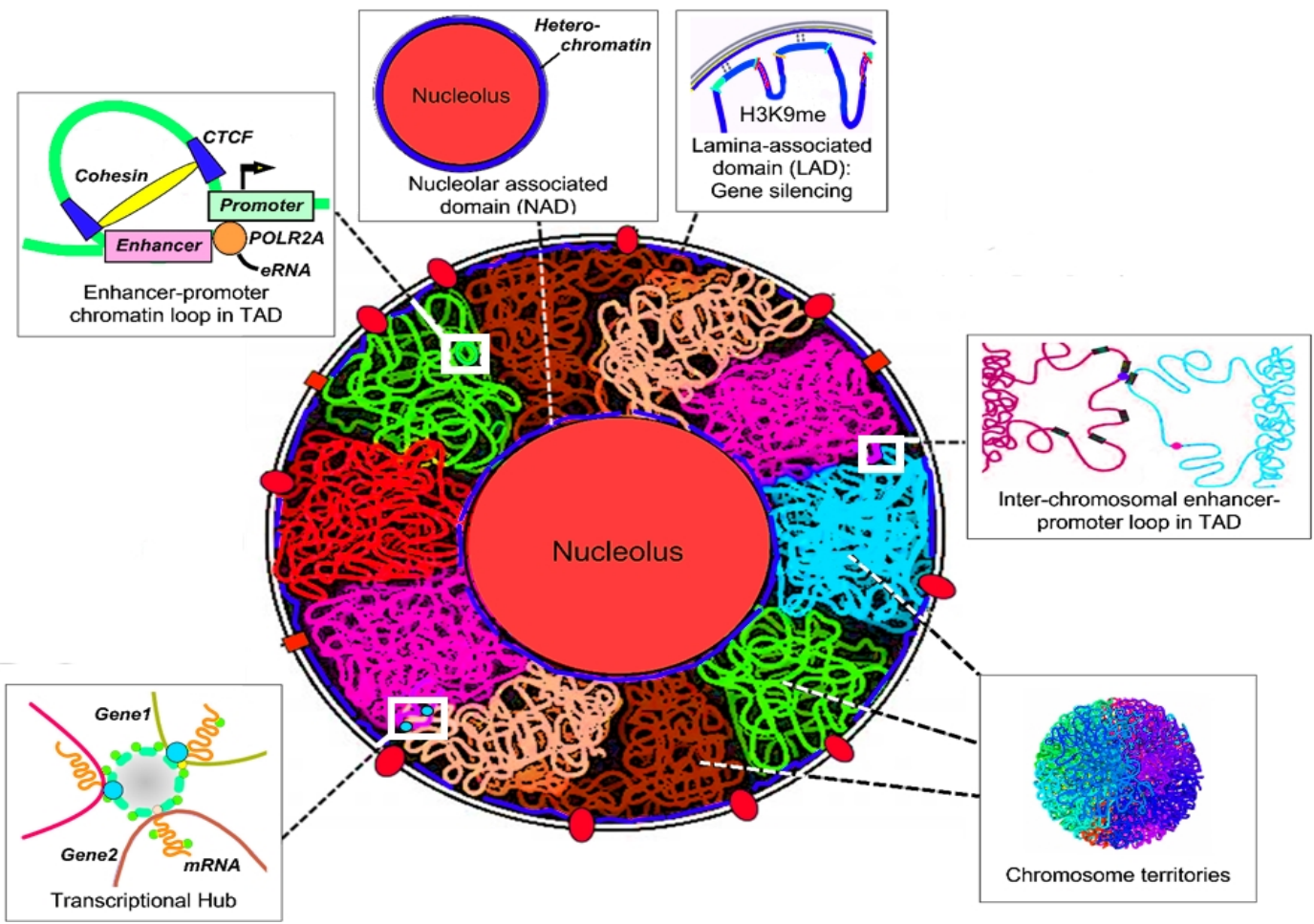

**Figure 5. Overview of different spatial regions that determine transcriptional state in the 3D Nucleome.** These include the fundamental unit of transcription, the topologically associated domains (TAD) and larger transcriptional hubs, embedded in chromosome territories (CTs). Active TADs tend to be located at the periphery of CTs. Enhancer-promoter looping in chromatin is an example of how distant regions in linear DNA sequence come together for regulation of gene expression. Heterochromatin located in perinucleolar and in lamina-associating domains (LADs), associated with histone marks such as H3K9me2/3, correlates with repression of gene expression. Inter-chromosomal interactions define one type of spatial *trans*-interaction. Adapted from [59], with permission from Elsevier.

The pharmacoepigenome can be defined as non-coding, regulatory regions of the genome that play an active role in the determination of medication efficacy, dose requirement, and adverse event (AE) profile [18]. The pharmacoepigenome contains gene regulatory elements such as enhancers and promoters. Since variation in non-coding regions in the human genome accounts for over 80% of the genetic contribution to disease risk, apart from the few known common single variants that cause Mendelian disorders [17], it is likely that variations in complex traits such as drug response and susceptibility to adverse drug events are also controlled by the noncoding genome [53]. Subsumed in this set is treatment-resistance, for example, regulatory non-coding elements power chemotherapeutic habituation to vemurafenib in 90% of melanoma patients [60], and resistance to 2 or more anti-epileptic drugs (AEDs) in approximately 30% of patients with epilepsy [61].

To infer attributes of non-coding SNPs regulatory and predict their impact on a phenotype, machine learning as well as probabilistic methods have been proposed. For example, to determine putative chromatin state annotation, applications based on Hidden Markov Models (HMMs) are still prevalent and are used to predict regulatory elements including promoters, enhancers, transcription start sites (TSS), gene bodies, etc., from SNPs found in genetic association studies [6]. These applications incorporate features such as histone marks that are characteristic of specific regulatory elements (see **Table 1** and **Figure 6**), localization of regulatory elements in open chromatin as indicated by DNase I hypersensitivity (DHS), disruption of transcription factor binding sites (TFBS), and quantitative trait loci (QTL).

| ELEMENT | HISTONES MODIFICATIONS |
|---|---|
| Transcriptional start site | H3K27ac, H3K4me3, open euchromatin |
| Active promoter | H3K27ac, H3K4me3, open euchromatin |
| Enhancer | H3K27ac, H3K4me1, H2A.Z, euchromatin |
| Gene body, towards 3' end | H3K79me2/3, H3K36me3 |
| Large introns | H3K9ac, H3K18ac, H3K36me1 |
| Translation | H3K4me1, H3K79me1 |
| Strong polycomb repression | H3K9me2,3, H3k27me2/me3 |

**Table 1. Table of associations between epigenomic elements and corresponding histone modifications.**

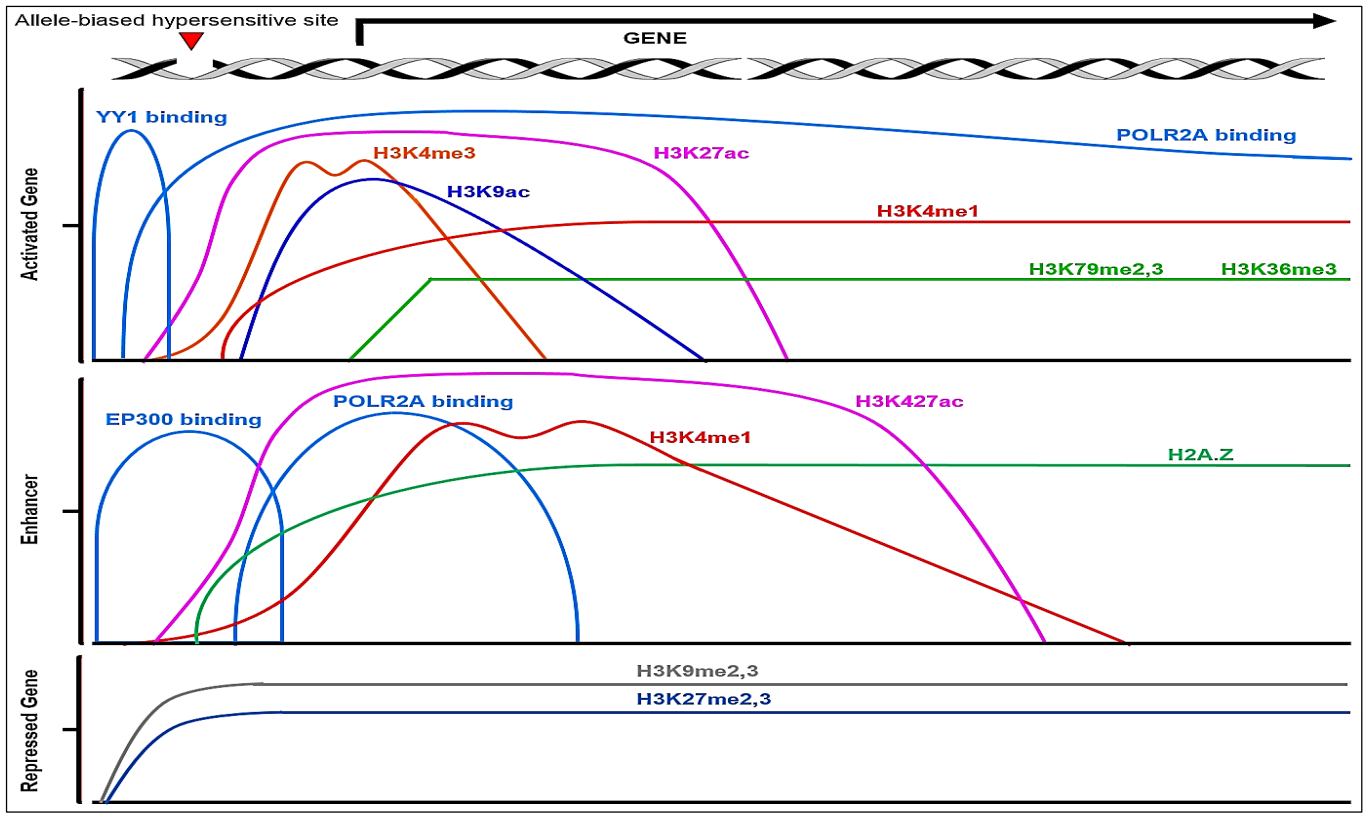

**Figure 6. Selected histone modifications used in current applications for prediction and classification of non-coding variants that detect gene regulatory elements.** Distribution of histone modifications and other characteristics of epigenomic regulatory elements; EP300: E1A Binding Protein P300; YY1: POLR2A: DNA-dependent RNA polymerase II; Ying Yang 1.

| SOFTWARE [REF] | SOURCE CODE | DESCRIPTION |
|---|---|---|
| **DeepSEA [47]** | http://deepsea.princeton.edu | Predicts the non-coding variant effects *de novo* from sequence by directly learning a regulatory sequence code from large-scale chromatin profile data, enabling prediction of chromatin effects of sequence alterations with single-nucleotide sensitivity. |
| **DeepBind [48]** | http://tools.genes.toronto.edu/deepbind/ | Predicts potential transcription factor binding sites (TFBS) and RNA binding protein (RBP) binding sites; both *in vitro* and *in vivo*, outperforming 26 previously tested algorithms. |
| **deepnet-rbp [62]** | https://github.com/thucombio/deepnet-rbp | Predicts RBP binding sites taking (predicted) RNA tertiary structural information into account. |
| **Basset [41]** | http://www.github.com/davek44/Basset | Predicts DNA accessibility, simultaneously learning the relevant sequence motifs and the regulatory logic with which they are combined to determine cell-specific DNA accessibility. Predictions for the change in accessibility between variant alleles are greater for Genome-wide Association Studies (GWAS) in SNPs that are likely to be causal relative to nearby SNPs in linkage disequilibrium with them. |
| **DanQ [63]** | https://github.com/uci-cbcl/DanQ | Uses the same features and data as the DeepSEA framework, outperforming DeepSEA for 97.6% of the targets. |
| **DeepChrome [64]** | https://github.com/QData/DeepChrome | Predicts gene expression from histone modification signals and enables the visualization of the combinatorial interactions among histone modifications via a novel optimization-based technique that generates feature pattern maps from the learned deep model. |
| **TFImpute [65]** | https://bitbucket.org/feeldead/tfimpute | Predicts cell-specific TF binding for TF-cell line combinations using a multi-task learning (MTL) setting to use information across TFs and cell lines. |
| **Rambutan [66]** | https://github.com/jmschrei/rambutan | Predicts Hi-C contacts at 1 kb resolution using nucleotide sequence and DNase I assay signal as inputs. Predicted contacts exhibit expected trends relative to histone modification Chromatin Immuno-Precipitation-Sequencing (ChIP-seq) data, replication timing measurements, and annotations of functional elements such as enhancers and promoters. |
| **CpGenie [67]** | https://github.com/gifford-lab/CpGenie/ | Produces allele-specific DNA methylation prediction with single-nucleotide sensitivity that enables accurate prediction of methylation quantitative trait loci (meQTL). Contributes to the prediction of functional non-coding variants, including Expression Quantitative Trait Loci (eQTL) and disease-associated mutations. |
| **DeepCpG [68]** | https://github.com/cangermueller/deepcpg | Identifies known and *de novo* sequence motifs that are predictive for DNA methylation levels or methylation variability, and to estimate the effect of single-nucleotide mutations. |
| **iDeep [69] and iDeepS [70]** | http://www.csbio.sjtu.edu.cn/bioinf/iDeep/ https://github.com/xypan1232/iDeepS | Predicts RBP binding sites by multimodal learning from multi-resource data, e.g. sequence, structure, domain specific features, and formats. Allows one to automatically capture the interpretable binding motifs for RBPs. |
| **FactorNet [71]** | https://github.com/uci-cbcl/FactorNet | Predicts TFBS by leveraging a variety of features, including genomic sequences, genome annotations, gene expression, and single-nucleotide resolution sequential signals, such as DNase I cleavage data. |
| **Basenji [72]** | https://github.com/calico/basenji | Predicts cell type-specific epigenetic and transcriptional profiles in large mammalian genomes from DNA sequence alone. Identifies promoters and distal regulatory elements and synthesizes their content to make effective gene expression predictions. Model predictions for the influence of genomic variants on gene expression that align well to causal variants underlying eQTLs in human populations; and can be useful for generating mechanistic hypotheses to enable GWAS loci fine mapping. |
| **Concise [73]** | https://github.com/gagneurlab/concise | Predicts RBP binding sites using a spline transformation-based neural network module to model distances from regulatory sequences to genomic landmarks. |
| **DeepATAC [74]** | https://github.com/hiranumn/deepatac | Predicts binding locations from both DNA sequence and chromatin accessibility as measured by ATAC-seq, outperforming current approaches including DeepSEA. |

**Table 2.** Examples of open-source deep learning software applications for the discovery of epigenomic regulatory interactions and variant annotation.

Several machine learning applications have been developed for predicting the impact of non-coding SNPs in GWAS on phenotypes; however, fewer than 40% of GWAS publications from 2015 utilized these tools [75]. **Table 2** lists some examples of deep learning software that score features, such as DHS for prioritization of regulatory function and protein annotation of chromatin loops, to predict functional enhancer-promoter interactions and drug-target inference.

Deep learning applications for detection of regulatory elements within the non-coding genome are beginning to emerge [7, 8, 76]. Most of existing applications are based on CNN architecture is trained either from *k*-mers [77, 78], or directly on genomic sequence data. For example, DeepSEA [47] is one of the first deep learning-based algorithmic frameworks for predicting the chromatin effects of sequence alterations with single nucleotide sensitivity. In addition, it is trained on diverse sets of chromatin profiles from ENCODE and Roadmap Epigenomics Consortium projects [17, 79]. DeepSEA can accurately predict the epigenetic state of a sequence, including transcription factors binding, DNase I sensitivities, and histone marks in multiple cell types. In addition, it can further utilize its capabilities to predict the chromatin effects of sequence variants and prioritize regulatory variants. In another example, the DeepBind algorithm was implemented based on a deep convolutional neural network to calculate the ability of nucleotide sequences to bind transcription factors and RNA-binding proteins in order to characterize the effects of single point mutations on binding properties in various diseases [48]. More recently, the Basset CNN model was used to predict DNA accessibility within non-coding regions [41]; and is intended to predict allele-bias in DNA accessibility, which is indicative of causal variants. DNase-Seq data from 164 cell types that had been mapped by ENCODE and the Roadmap Epigenomics Consortium was used to create Basset. The Basset CNN learned both protein-DNA binding motif information, as well as the underlying regulatory knowledge that determines cell-specific DNA accessibility. In the analysis of GWAS SNPs that were determined to be casual autoimmune variants, Basset demonstrated that it could discriminate causal from non-casual SNPs in high Linkage Disequilibrium (LD). In contrast to inference of regulatory elements using annotation based on pre-defined feature sets, models such as DeepSEA and Basset do not take handcrafted, preprocessed features. Instead, they adaptively learn them from raw sequence data during the training phase. This, combined with high expressive power, allows deep learning to outperform traditional machine learning models. More accurate prediction of non-coding variants and their functional annotations with deep learning methods promises to enable better understanding of pharmacoepigenomic variation and more accurate prediction of drug response and adverse events (AEs).

Other recent applications of deep learning models to prediction of regulatory elements and their interactions with the state-of-the-art performance include enhancer prediction [80-82]; classification of gene expression using histone modification data as input [64]; prediction of DNA methylation states from DNA sequence and incomplete methylation profiles in single cells [68]; prediction of enhancer-promoter interactions from genomic sequence [83]; prediction of DNA-binding residues in proteins [84]; global transcription start prediction [85]; and improved prediction of the impact of non-coding variants on DNA methylation [67, 85]. In 2016, Google and Verily Life Sciences published a pre-print describing "DeepVariant" – a deep learning-based universal SNP and small indel variant caller that won the "highest performance" award for the SNPs Administration-sponsored variant calling "Truth Challenge" in May 2016 [86]. Recently, an updated, open-source version of DeepVariant has been further evaluated on a diverse set of additional tests by DNAnexus [87]. These tests showed that application of a general deep learning framework exceeded the accuracy of traditional methods for SNP and indel calling that has been developed over the last decade. Deep neural networks also demonstrated the ability to outperform conventional machine learning techniques in SNP–SNP interaction prediction [88, 89].

## 3.2 The application of artificial intelligence for patient stratification

The FDA provided guidance in 2013 [90] that pharmacogenomic testing should be used in early-phase clinical trials for the identification of populations, cohorts, and individuals "that should receive lower or higher doses of a drug, or longer titration intervals, based on genetic effects on drug exposure, dose-response, early effectiveness, and/or common adverse reactions." Although this approach has not been widely adopted by pharmaceutical companies, in part, for fear of reducing its potential market size and a lack of available large genomic data resources, applications of AI methods for patient stratification using clinical data are beginning to see usage and adoption [91]. The industry seems to be moving towards the direction of developing proprietary machine learning algorithms to stratify patients using both unstructured and structured data obtained from the client's electronic health records (EHRs), which can be applied to both clinical research in academia as well as clinical trials in pharmaceutical research. By leveraging genomic data, and information from HapMap and 1000 Genomes Project [92], ethnicity stratification can be performed on population, cohort, and other levels. This can be further extended by using EHRs for improved patient stratification, potentially leading to more precise risk models designed to advance clinical and translational research. Indeed, AI continues to rapidly advance in the biomedical research domain, with examples of deep learning-based methods recently gaining FDA clearance for the clinical usage with cardiac MRIs [16].

Patient stratification involves the complex integration of heterogeneous biomedical, demographic, and socio-metric data to categorize patients into subpopulations for design of clinical trials and clinical practice. In this context, data mining of electronic health records has been proposed as an efficient relevance-based method to potentially identify eligible patients for clinical trials [93]. Despite not being designed for usage in research, substantial amounts of data within EHRs, such as surrogate disease phenotypes imputed from ICD codes, have effectively been proven for use through several notable studies in GWAS and Phenome Wide Association Studies (PheWAS) analysis [94]. Furthermore, studies on EHR-linked DNA bio-repositories have successfully shown that integration of such pharmacogenomics and sociometric data can be useful in predictive modeling for optimizing dosage and reducing dosing error [94]. By using clinically available information, such as age, gender, and education, health-care providers and clinical researchers can identify better treatment options and patient responses to maximize efficacy and cost-effectiveness [94, 95], as shown in **Figure 1**.

However, there are several challenges associated with the effective integration of EHR data for pharmacogenomics applications. For example, due to high dimensionality of the EHR data structure, noise, heterogeneity, sparseness, incompleteness, random error, and systematic biases [96], extraction of relevant clinical phenotypes may require extensive feature engineering and advanced computational models beyond traditional machine learning methodologies. Ongoing research in this domain and recent advances in deep learning demonstrate the potential of deep learning to overcome these challenges and learn patient data representations that are useful for treatment response and outcome prediction [91]. Recent applications in this area include extraction of general-purpose patient representations from EHRs, often performed with generative models trained either on static or temporal data [91]. These models are capable of uncovering patterns in sparse, complex, heterogeneous datasets and producing surrogate imputed patient phenotypes. For example, there are both unsupervised, e.g. Deep Patient [96], and semi-supervised, e.g. Denoising Auto-encoder for Phenotype Stratification (DAPS) [40], models that rely on a stacked auto-encoder network architecture to model EHR data to derive patient representations that are predictive of final diagnosis, patient risk level, and outcome (e.g. mortality, re-admission) (**Table 3)**. As generative deep model development progresses quickly, applications of novel architectures, such as Generative Adversarial Networks (GANs), to EHR data are starting to emerge, demonstrating improved performance for the disease prediction [97] and risk prediction given treatment [98].

Overall, the promise of integrating pharmacogenomics with data-driven EHR analysis of population, cohort, and individual patient data already shows usefulness for patient stratification and prediction of treatment response. A quickly growing body of work in the field demonstrates a great interest in applying deep networks to these problems [91], which allows one to learn from heterogeneous EHR data, extract temporal patterns, impute missing data, and predict clinical outcomes and optimal treatment strategies while outperforming conventional machine leaning methods.

### 3.3 Deep learning for temporal patient data

Due to the longitudinal nature of EHR data, many applications employ deep network architectures that are capable of extraction of temporal patterns from it, such as recurrent neural networks (RNNs), Long Short-Term Memory networks (LSTMs), etc. [91]. These networks are used for mapping patient trajectories with temporal predictions of clinical outcomes, outperforming conventional machine learning methods that typically require a single "snapshot" in time, and are not as robust for longitudinal modeling [99-101]. Several methods have been proposed to deal with the complex nature of longitudinal EHR data, specifically because of temporality from clinical records. To account for possible interventions and predict optimal treatment strategy, deep learning approaches were shown to be efficient when combined with reinforcement learning. For example, Kale *et al*. demonstrated how this type of deep model can be used for discovery and analysis of causal phenotypes from clinical time series data [102]. Deep neural networks trained on EHR data with temporally dependent constraints and outputs have also been proposed to predict 3-12-month mortality of patients receiving improved palliative care [46]. Additional deep reinforcement learning models have been used to learn an optimal heparin dosing policy from sample dosing trials. Their associated outcomes having been predicted from the publicly available MIMIC II intensive care unit database [103].

Looking forward, we also envision the incorporation of data from mobile devices and wearable sensors for measuring phenotypic markers and stratification of patients by these phenotypes. This type of continuously collected data allows researchers access to large-scale deep phenotyping of the human population, and to better assess patients' prognosis by analyzing their real-time data. Rajpurkar et al. developed a 34-layer convolutional neural network which exceeded the performance of board certified cardiologists in detecting a wide range of heart arrhythmias from electrocardiograms recorded with a single-lead wearable monitor [104]. Apple's ResearchKit open-source framework enables access to enrolled patients' heart rate, accelerometer, and other mobile sensor data [105]. For example, the approach utilizing deep convolutional neural networks for feature extraction from accelerometry and gyroscope iPhone data has recently won the Parkinson's Disease Digital Biomarker (PDDM) DREAM challenge, an open crowd-sourced research project designed to benchmark the use of remote sensors to diagnose and track Parkinson's disease [106]. Similar studies with recurrent neural networks have also shown to be successful in classification of patients with bipolar disorder using NLP and accelerometer data collected from a patient's mobile device [107]. Although data from wearable sensors isn't yet considered to be a part of a patient's electronic health record, this data has shown to be robust and usable with deep learning methods and will certainly contribute to the modernization of patient stratification.

| SOFTWARE [REF] | SOURCE CODE | DESCRIPTION |
|---|---|---|
| **Deep Patient [96]** | https://github.com/staplet14/DeepPatient <br> https://github.com/natoromano/deep-patient | Learns a general-purpose patient representation from EHR data in unsupervised manner that is broadly predictive of health status as assessed by predicting the probability of patients to develop various diseases. Results significantly outperforming those achieved using representations based on raw EHR data and alternative feature learning strategies. |
| **DeepCare [101]** | https://github.com/trangptm/DeepCare | Predicts unplanned readmission and high-risk patients for the diabetes and mental health patient cohorts using EHR data including diagnosis, procedure, and medication codes. Outperforms SVM, random forests, "plain" RNN, and LSTM with logistic regression. |
| **Doctor AI [99]** | https://github.com/mp2893/doctorai | Implements a generic predictive model that covers observed medical conditions and medication uses from longitudinal time-stamped EHR data. Performance was judged on classification of the final diagnosis (aggregated to 1183 unique ICD9 codes) and prediction of medical order (grouped into 595 unique GPI codes). |
| **MIMIC trajectories [100]** | https://github.com/EpistasisLab/MIMIC_trajectories | Learns meaningful representations from a longitudinal sequence of a patient's interactions with the health care system (care events) in both unsupervised and supervised settings that are shown to be useful for patient survival prediction. |

**Table 3**. Examples of recently published research and open-source deep learning software for applications of artificial intelligence (AI) in patient stratification and patient care coordination.

### 3.4 Pharmacological applications for drug and target discovery, repurposing, and interaction

Although the non-coding human genome represents the new source for drug targets and genetic variation discovery, so far, most approaches to "epigenetic" drug discovery have focused on post-translational modification of histone proteins and DNA through enzymes ("writers" and "erasers"), and the recognition of these changes by adaptor proteins ("readers"). There are now hundreds of identified chromatin remodeling proteins that aggregate into larger protein complexes, and exert complex functions such as chromatin-mediated neuroplasticity and neurogenesis in the human CNS. These chromatin remodeling proteins were first examined in the context of developmental decisions about cell fate, and in the adult, potential druggable targets were thought to consist of histone demethylases (e.g., KDM1A), histone methyltransferases such as EZH2, and bromodomain-containing proteins, that were thought to be the only 'readers' of the histone code. On closer inspection, these turned out to be super-families of related proteins, and there exist many other proteins that act as chromatin remodelers **(Figure 5)**.

The realization that the human epigenome operates the fundamental regulatory machinery of transcription, many new druggable targets can be discovered that are not "writers", "erasers" or "readers." Fundamental to this realization was the recognition that the linear genetic sequence was only the beginning, as the important mechanisms of gene regulation operate in the spatial and temporal dynamics between regulatory elements such as super-enhancers, enhancers and promoters along with the target genes they regulate. Additionally, several important characteristics of causal variants in GWAS have emerged, including properties such as allele-specific bias, location of gene variants in euchromatin, and histone marks that are associated with genome elements that help define their function. Critical to transcriptional regulatory circuits was the realization that transcription factors are key drivers of phenotype, and they can be classified in a hierarchal manner. In addition, master transcription factors are controlled by super-enhancers for determination of cell-specific gene regulation and identity [108].

Ivanov et al. [109] first recognized the complexity of the molecular physiology responsible for regulation of ADME genes, including DNA methylation and hydroxymethylation, various histone modifications, miRNAs, and lncRNAs. Since a xenobiotic substance that alters any of the myriad of enzymes and small RNAs involved in ADME gene regulation represents a novel therapeutic candidate, they emphasized the importance of "pharmacoepigenomics" in drug discovery. Since 2012, our understanding of the druggable epigenome has increased exponentially, providing thousands of new druggable targets **(Figures 5, 6)**.

Although the human epigenome has yielded insight into pharmacogenomic regulatory mechanisms, translation of this wealth of data into drug discovery will not be trivial. Currently, although the single most valuable approach for detection of *prospective* druggable targets in the human epigenome is the application of deep learning methods for candidate identification **(Table 2)**, the ability to test compound/drug – molecule pairs is hobbled by protracted preclinical screening in animal models [29, 32]. The cost and time incurred by the brute force screening of thousands of small compounds for novel epigenome drug targets in animal models is a daunting challenge. Simulation of the mechanism of candidate drugs action in populations of "virtual humans", which accurately represent the molecular physiology of actual humans is within reach [110], but is an under-funded application domain in biomedical informatics and computer science. Translation will require innovation in domains that have traditionally resisted change, including adoption of adaptive clinical trial design, transformation of federal regulatory governance, and broad adoption in preclinical pharmaceutical research of genomics-based science and *in silico* strategies that have been shown to be effective in predicting the clinical success of drug targets [111].

Deep learning applications in drug discovery and repurposing are starting to emerge, and already show high potential in many tasks including virtual screening, ADME/Tox properties prediction, novel drug-target interactions prediction [8, 28-32]. Generative deep models are also used for *de novo* design molecules with desired chemical properties [112-114]. In **Table 4** we show those applications with publicly available open-source implementations.

### 3.5 Deep Learning and toxicology

The Tox21 Data Challenge has been the largest effort of the scientific community to compare computational methods for toxicity prediction [115]. This challenge comprised 12,000 environmental chemicals and drugs that were measured for 12 different toxic effects by specifically designed assays. In this challenge, deep learning based DeepTox model had the highest performance of all computational methods winning the grand challenge, the nuclear receptor panel, the stress response panel, and six single assays [115]. DeepTox also demonstrated the benefit of using a multi-task network, which outperformed a single-task counterpart in 10 out of 12 assays. Further studies also suggest that multi-task deep networks show superior performance on a broad range of drug discovery datasets [31, 116].

DeepAOT family of deep architectures for the compound acute oral toxicity prediction is based on molecular graph encoding convolutional neural networks [117]. These models implement regression, multi-classification, and multi-task networks that outperformed previously reported models for this task. Interpretation of these models was performed by exploration of networks' internal features (referred to as deep fingerprints) that were highly correlated with topological structure-based fingerprints. Furthermore, one toxicity-related feature of each deep fingerprint was tracked back to the atomic level and the highlighted toxicity fragments were then compared with structural alerts for toxic chemicals and compounds with potential adverse reactions from ToxAlerts database [118]. Consistent results suggested that DeepAOT models could infer acute oral toxicity related toxic fragments from just the information on molecular shape and atomic bonds. Moreover, this model architecture is not limited to acute oral toxicity, and it could be applied for studying other end points induced by compounds in complex systems [117].

| SOFTWARE [REF] | SOURCE CODE | DESCRIPTION |
|---|---|---|
| **DeepChem** [119] | https://deepchem.io/ | Implements low data learning method based on a novel iterative refinement long short-term memory architecture combined with graph convolutional neural networks to learn of meaningful distance metrics over small-molecules. On the Tox21 and SIDER collections, one-shot learning methods strongly dominate simpler machine learning baselines, indicating the potential for strong performance on small biological data sets. |
| **DeepDTIs** [120] | https://github.com/Bjoux2/DeepDTIs_DBN | Training drug-target space extracted from DrugBank consisted of 1412 drugs and 1520 targets. Experimental drug-target pairs for testing were derived from DrugBank as well and consisted of 2528 targets and 4383 experimental drugs. DeepDTI gained the best performance in multiple performance metrics as compared to the Bernoulli Naive Bayesian, Decision Tree, and Random Forest classifiers. In addition, DeepDTI showed potential to predict whether a new drug targets to some existed targets, or whether a new target is interacting with some existed drugs. |
| **DeepSynergy** [121] | https://github.com/KristinaPreuer/DeepSynergy | Predicts synergistic drug combinations for cancer therapy by learning from chemical properties of the drugs and gene expression profiles of specific cancer cell lines. DeepSynergy significantly outperformed the other methods with an improvement of 7.2% over the second-best method at the prediction of novel drug combinations within the space of explored drugs and cell lines. Applying DeepSynergy for classification of these novel drug combinations resulted in a high predictive performance of an AUC of 0.90. |

**Table 4.** Examples of open-source deep learning software applications for pharmacological applications.

### 3.6 Other pharmacogenomic applications

Applications of deep learning recently demonstrate state-of-the-art performance for predicting cell phenotypes from transcriptomics data [122], drug response in cancer [123], seizure-inducing side effects of preclinical drugs [124], patient survival from multi-omic data [38], drug-induced liver injury prediction [62], and classifying genomic variants into adverse drug reactions [125].

## 4 Future perspectives

### 4.1 Impact on basic research in biology and pharmacology

We anticipate that as larger and more heterogeneous pharmacogenomic datasets become available in coming years, the predictive power of AI, and specifically deep learning models, will increase. Abundance of various types of data not only will enable more effective data-driven mining and discovery of important variants and markers, but allow as well for deeper investigation of corresponding interaction mechanisms by systematically considering underlying biological processes at different scales and biological data modalities. As already noted, discovery and characterization of non-coding regulatory elements in 4D Nucleome already are becoming a major topic of pharmacogenomic studies. Further investigations will continue this trend, with a focus on causal relationship between elements of interest and increasing model specificity and predictive power originating from multi-omics and pathway studies.

Increasing amounts of patient-specific data such as EHRs, environmental data, and demographics, combined with pharmacogenomic targets and pharmacological knowledge bases will allow patient stratification into treatment groups with specificity at the population, cohort, and individual levels. Advanced machine learning models such as deep learning will allow the researcher to jointly learn multiple objectives from heterogeneous, multimodal data and predict, for example: novel variants, their effects and functions, drug adverse event (AE) risk estimation, treatment and dosage recommendation, and other pharmacological outputs (**Figure 7**). Given the growing amount of these data, AI methods, including deep learning, often demonstrate the best performance in addressing relevant methodological challenges [8]. However, as discussed above, applications of any machine learning algorithm, including deep learning, require careful selection of controls, training sets, and appropriate validation schemes and metrics--and all of these should be combined with domain expertise to fully realize the potential of AI and deep learning.

### 4.2 Industrial perspectives

Pharmaceutical companies were quick to recognize the potential application of AI methods such as deep learning for drug discovery and development. Market forecasts emphasized in 2017 that the "full potential healthcare service cost savings of AI-enabled initiatives would be $300 billion a year in the United States, or about 0.7 percent of GDP" [126], and "big pharma, biotech, CROs and research institutes will spend $390M US on deep learning for drug discovery, including products, services and internal projects worldwide, and this market will grow to $1.25B by 2024" [127]. This led to a rush of investment into start-ups intending to offer AI consulting services and products offered to pharmaceutical companies, with a similarity to the rise of new companies offering companion diagnostics in the early part of the 21st century, which did not reach profitability in an entrepreneurial timeframe. There has been a reluctance of the large pharmaceutical companies, whose culture is often monolithic and conservative, to embrace innovation in the absence of pragmatic demonstration that AI methods would lead to success in clinical trials. More importantly, big pharma quickly realized that access to large quantities of genotyped patient data was a major priority for patient stratification and other applications, but in the U.S., security concerns has limited partnerships among health-care data owners—hospitals, insurers, and drug makers. This has recently led big pharma to take the position of moving quickly to organize and curate their own datasets for internal use, and form partnerships with national bio-banks and other entities for external forces of data, while engaging in watchful waiting for more realistic demonstrations of the practicality of deep learning for drug discovery and development before large investments are made. During this time, the industry as a whole has remained in vigorous discussions with various stakeholders about the potential and limitations of such methods. Now that there is some amount of critical evaluation of AI applications in drug discovery [128], the pharmaceutical industry is poised to embrace omics-driven drug discovery, phenotype-driven drug discovery, and stratification in clinical trials on a massive scale as soon as convincing validation emerges from current efforts.

**4.3 Open science considerations**

A culture of data, code, and model sharing promises to speed advances of deep learning applications in pharmacogenomics. The sharing of high quality, labeled datasets will be especially valuable; however, a clear asymmetry exists with government-sponsored academic researchers directed to share, while researchers in industry are often prohibited from sharing code, data, and results due to proprietary and intellectual property protections. Availability of open-source solutions for the discovery of epigenomic regulatory interactions and variant annotation (**Table 2**) compared to those in patient stratification (**Table 3**) and pharmacological applications (**Table 4**) shows that the potentially translational character and patient privacy aspects of the study often prevents its details, data, and implementation from being open to public. However, this situation has already changing in the machine learning and deep learning communities, that has witnessed acceleration of progress via public-posting of various datasets for benchmarking and software tools, including those developed and used in the industrial setting. Moreover, researchers who invest time to pre-process datasets to be suitable for deep learning can make the pre-processing code (e.g. Basset [38]) and cleaned data publicly available to catalyze further research. Code-sharing and open-source licensing is essential for continued progress in this domain. Because many deep learning models are often built using one of several popular software frameworks, it is also possible to directly share trained predictive models. A pre-trained neural network can be quickly fine-tuned on new data and used in transfer learning, as discussed above. It is possible for models to be trained on competitive proprietary data without the release of such data, and a consortium model of joint training on proprietary data from multiple sources has been contemplated within the industry.

## 4.4 Conclusions

Deep learning models will be further improved to address current limitations such as training time, interpretability of results, and requirements to training set size. Transfer learning, that involves training deep learning models on one type of data and adaptation of learn representation to another type, will be commonly used in experiments where data collection remains expensive. Models such as deep neural networks will be adapted to learn joint data representation from various omics data types, which will allow combining information from different experiments to provide better-informed predictions. As dataset sizes increase, we will see more of semi- and un-supervised machine learning applications, including generative models that will be able to produce or suggest new and testable biological hypotheses, such as potential novel pharmacogenomic markers (PK, PD, and ADME), and also drug targets, based on information extracted from multi-modal omics data.

Several decades from now, one can imagine the situation when machine learning and artificial intelligence-based systems will shift their focus from "prediction" to "prescription", i.e. will not only provide insights, but will also provide recommendations for further action. Such changes not only have the potential to revolutionize pharmacogenomics and pharmaceutical research more broadly, but also will likely provide a wide impact on biological sciences, and on health, in general.

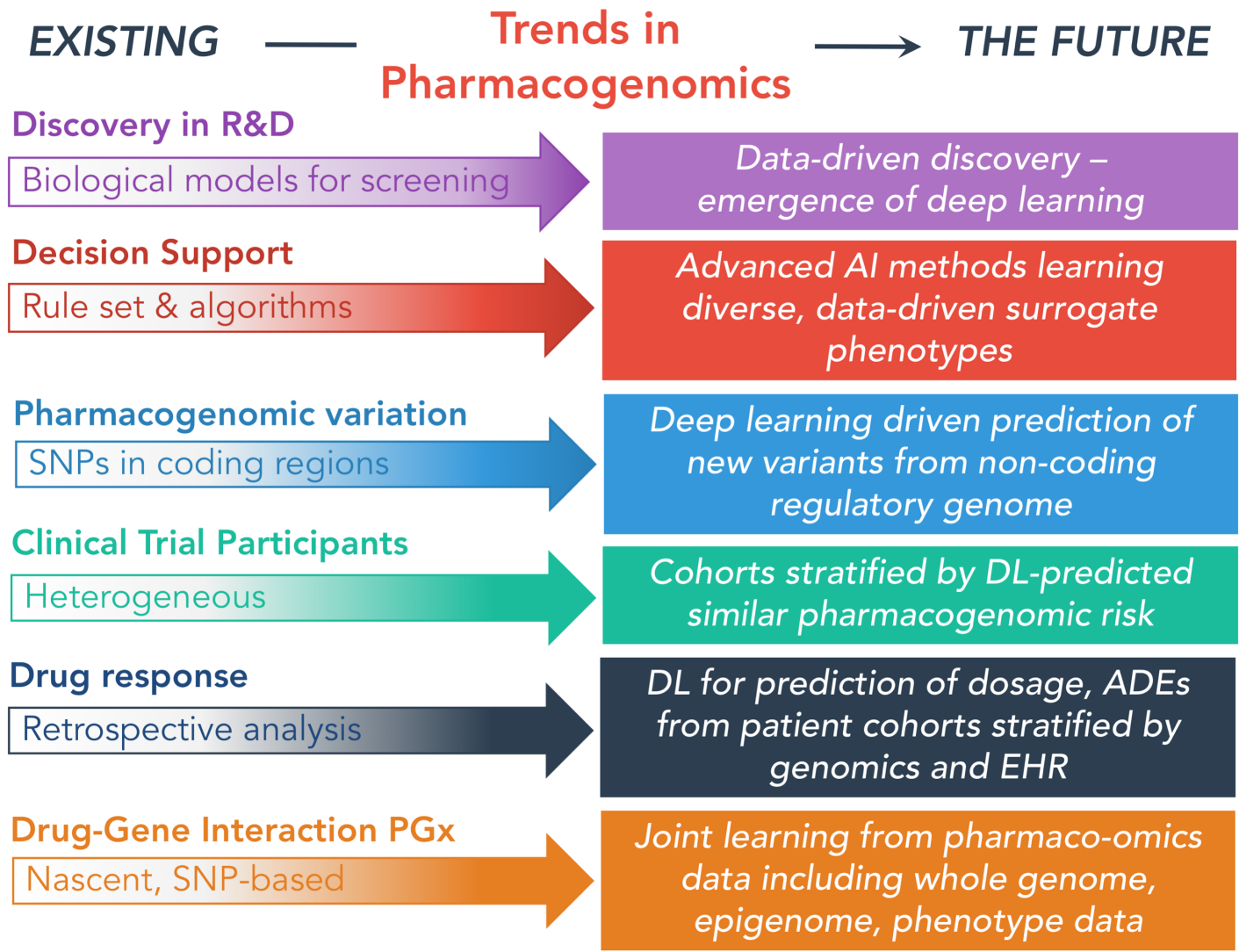

**Figure 7. Future trends in pharmacogenomics.** We anticipate that as larger and more heterogeneous pharmacogenomic datasets become available in coming years, the predictive power of deep learning models will increase.

| Executive summary |
|---|
| *Pharmacogenomics has promising applications in drug discovery and development, and medication optimization* |
| <ul><li>Pharmacogenomic studies have established the importance of drug-genome interactions.</li><li>Pharmacogenomics offers promise for applications such as medication optimization for patients based on genotype in diagnostic testing, value as a companion diagnostic (CDx), and drug discovery and development.</li><li>The non-coding regulatory genome is the current domain for the discovery of new genomic variants.</li></ul> |
| *Brief overview of machine learning methodology* |

- **Machine learning methods have demonstrated the ability to identify novel regulatory variants located in non-coding domains that can inform pharmacogenomic response, prediction of drug-genome interactions, and extraction of pharmacogenomic phenotype from clinical data, and drug discovery.**
- **The predictive power of machine learning is realized mostly when it is combined with prior domain knowledge, such as gene networks and pathways.**
- **Usage of traditional machine learning models is challenged by rapid growth of data volume, a need for combination of heterogeneous datasets from different experiments, and is highly resource intensive in its pre-processing and feature handcrafting application(s).**

*Deep learning takes advantage of big data via representation learning*

- **Deep learning is subset of machine learning models composed of multiple processing layers to learn representations of data with multiple levels of abstraction, which eliminates the feature extraction step. These models improved the state-of-the-art in many machine learning tasks, including several examples in genomics and in and drug discovery.**
- **Applications of deep learning in pharmacogenomics have started to emerge, but they are still in its' infancy.**
- **Deep learning models often require relatively large training sets, architecture design, and careful choice of validation techniques to prevent overfitting.**

*Identification of regulatory pharmacogenomic variants and drug target discovery using deep learning*

- **Significant molecular variation which accounts for human differences in medication response and adverse events may be based in the intricate organization of the 4D spatial genome (or 4D Nucleome), which drives a need for deeper investigation of nuclear zones of transcriptional regulation.**
- **Pharmacoepigenomic datasets contain information about gene regulatory elements such as promoters and enhancers, histone marks, disruption of transcription factor binding sites (TFBS) and quantitative trait loci (QTL). They are used in training of various machine learning models to infer regulatory attributes of non-coding SNPs, including chromatin state annotation, promoters, enhancers, transcription start sites, gene bodies, etc.**
- **Deep learning models already have demonstrated state-of-the-art performance in several tasks such as predicting DNA accessibility within noncoding regions, potential TF and RNA binding sites, and gene expression from histone modifications.**
- **Realization, that many of genetic differences in drug response can be found via analysis of noncoding regulatory genome together with modern gene editing techniques opens great opportunities for applications of machine and deep learning to predict the key regulatory variants that impact drug response and induce adverse drug events.**
- **Applications of deep learning for phenotype extraction from medical records and other patient data, including temporal, show the potential usefulness for pharmacogenomic patient stratification, individualized treatment outcome prediction and medication optimization.**
- **The fact, that human epigenome operates the fundamental regulatory machinery of transcription in the spatial and temporal dynamics, suggest that discoveries of many new druggable targets that were not in the focus of traditional "epigenetic" drug discovery, are potentially realizable.**

- **Deep learning can be a lead force in pharmacoepigenomics-based drug discovery, combining candidate prediction with virtual screening, *in silico* drug repurposing and evaluation.**